\documentclass[sigconf]{acmart}

\AtBeginDocument{%
\providecommand\BibTeX{{%
\normalfont B\kern-0.5em{\scshape i\kern-0.25em b}\kern-0.8em\TeX}}}

\usepackage{amsmath}
\usepackage{booktabs}
\usepackage[capitalise]{cleveref}
\usepackage{dirtytalk}
\usepackage{enumitem}
\usepackage{pifont}
\usepackage{ifthen}
\PassOptionsToPackage{table}{xcolor}

\colorlet{fg}{gray}

\usepackage{algorithm2e}
\RestyleAlgo{ruled}

\usepackage{listings}

\usepackage{subcaption}
\AtBeginDocument{\DeclareCaptionSubType{lstlisting}}
\crefname{sublstlisting}{listing}{listings}
\Crefname{sublstlisting}{Listing}{Listings}

\usepackage{makecell}
\usepackage{multirow}

\usepackage[nolist]{acronym}
\usepackage[group-minimum-digits=3]{siunitx} 

\renewcommand{\see}[1]{(\cf~\cref{#1})}

\makeatletter
\newcommand{\source}{\@ifnextchar\bgroup{\@source}{\@source{}}}
\newcommand{\@source}[1]{\textcolor{red}{(source\if\relax\detokenize{#1}\relax\else: #1\fi)}}
\makeatother

\def\eg{\emph{e.g.}}

\def\cf{\emph{cf.}}

\def\rq#1{\emph{RQ#1}}
\def\c#1{\emph{C#1}}
\def\r#1{\emph{R#1}}

\newcommand{\abs}[1]{\ensuremath{\left\lvert #1 \right\rvert}}
\newcommand{\mymax}[1]{\ensuremath{\left\lceil #1 \right\rceil}}
\newcommand{\mymin}[1]{\ensuremath{\left\lfloor #1 \right\rfloor}}
\newcommand{\avg}[1]{\ensuremath{\overline{#1}}}
\newcommand{\median}[1]{\ensuremath{\widetilde{#1}}}

\newcommand{\myquote}[1]{\fbox{\begin{minipage}{.9\linewidth}\say{\emph{#1}}\end{minipage}}}

\newcommand{\threatreport}{\textsc{ThreatReport}}
\newcommand{\cysecalert}{\textsc{CySecAlert}}

\begin{acronym}
	\acro{bow}[BoW]{Bag of Words}
	\acro{bsi}[BSI]{Bundesamt für Sicherheit in der Informationstechnologie}
	\acro{cert}[CERT]{Computer Emergency Response Team}
	\acro{cti}[CTI]{Cyber Threat Intelligence}
	\acro{dbscan}[DBSCAN]{Density-Based Spatial Clustering of Applications with Noise}
	\acro{dbstream}[DBSTREAM]{Density-Based Stream Clustering}
	\acro{ioc}[IoC]{Indicator of Compromise}
	\acroplural{ioc}[IoCs]{Indicators of Compromise}
	\acro{mc}[MC]{Micro-Cluster}
	\acroplural{mc}[MC]{Micro-Cluster}
	\acro{cve}[CVE]{common vulnerability and exposure}
	\acro{nvd}[NVD]{National Vulnerability Database}
	\acro{ml}[ML]{machine learning}
	\acroplural{cve}[CVEs]{common vulnerabilities and exposures}
	\acro{nist}[NIST]{National Institute of Standards and Technology}
	\acro{nltk}[NLTK]{Natural Language Toolkit}
	\acro{optics}[OPTICS]{Ordering Points To Identify the Clustering Structure}
	\acro{osint}[OSINT]{Open Source Intelligence}
	\acro{tfidf}[TF-IDF]{Term Frequency-Inverse Document Frequency}
	\acro{tip}[TIP]{Threat Intelligence Platform}
	\acro{t-sne}[t-SNE]{t-Distributed Stochastic Neighbor Embedding}
	\acro{url}[URL]{Uniform Resource Locator}
	\acro{bert}[BERT]{Bidirectional Encoder Representations from Transformers}
	\acro{sbert}[SBERT]{Sentence-BERT}
	\acro{csa}[CSA]{cyber situational awareness}
	\acro{nn}[NN]{neural network}
	\acro{oov}[OoV]{out-of-vocabulary}
	\acro{sbr}[SBR]{security bug report}
	\acro{soc}[SOC]{Security Operations Center}
	\acro{llm}[LLM]{large language model}
	\acro{mse}[MSE]{Microsoft Exchange}
\end{acronym}

\begin{document}

\title[Reducing Information Overload]{Reducing Information Overload:\\ Because Even Security Experts Need to Blink}

\author{Philipp Kuehn}
\email{kuehn@peasec.tu-darmstadt.de}
\orcid{0000-0002-1739-876X}
\authornote{Corresponding author}
\affiliation{%
	\institution{Science and Technology for Peace and Security (PEASEC), Technical University of Darmstadt}
	\city{Darmstadt}
	\country{Germany}}

\author{Markus Bayer}
\email{bayer@peasec.tu-darmstadt.de}
\orcid{0000-0002-2040-5609}
\affiliation{%
	\institution{Science and Technology for Peace and Security (PEASEC), Technical University of Darmstadt}
	\city{Darmstadt}
	\country{Germany}}

\author{Tobias Frey}
\email{tobiasjonathan.frey@stud.tu-darmstadt.de}
\affiliation{%
	\institution{Science and Technology for Peace and Security (PEASEC), Technical University of Darmstadt}
	\city{Darmstadt}
	\country{Germany}}

\author{Moritz Kerk}
\email{kerkmoritz1@gmail.com}
\affiliation{%
	\institution{Science and Technology for Peace and Security (PEASEC), Technical University of Darmstadt}
	\city{Darmstadt}
	\country{Germany}}

\author{Christian Reuter}
\email{reuter@peasec.tu-darmstadt.de}
\orcid{0000-0003-1920-038X}
\affiliation{%
	\institution{Science and Technology for Peace and Security (PEASEC), Technical University of Darmstadt}
	\city{Darmstadt}
	\country{Germany}}

\settopmatter{printacmref=false}
\renewcommand\footnotetextcopyrightpermission[1]{}
\pagestyle{plain}

\ccsdesc[500]{Computing methodologies~Cluster analysis}
\ccsdesc[500]{Computing methodologies~Supervised learning}
\ccsdesc[100]{Security and privacy~Usability in security and privacy}

\begin{abstract}
	\acp{cert} face increasing challenges processing the growing volume of security-related information.
	Daily manual analysis of threat reports, security advisories, and vulnerability announcements leads to information overload, contributing to burnout and attrition among security professionals.
	This work evaluates \num{196} combinations of clustering algorithms and embedding models across five security-related datasets to identify optimal approaches for automated information consolidation.
	We demonstrate that clustering can reduce information processing requirements by over \num{90}~\% while maintaining semantic coherence, with deep clustering achieving homogeneity of \num{.88} for \ac{sbr} and partition-based clustering reaching \num{.51} for advisory data.
	Our solution requires minimal configuration, preserves all data points, and processes new information within five minutes on consumer hardware.
	The findings suggest that clustering approaches can significantly enhance \ac{cert} operational efficiency, potentially saving over \num{3750} work hours annually per analyst while maintaining analytical integrity.
	However, complex threat reports require careful parameter tuning to achieve acceptable performance, indicating areas for future optimization.
	The code is made available at \url{https://github.com/PEASEC/reducing-information-overload}.
\end{abstract}


\keywords{Clustering, Security, Machine Learning, \aclp*{cert}}

\maketitle
\acresetall

\section{Introduction}
\label{clustering:sec:intro}

The cybersecurity threat landscape continuously evolves, with attackers deploying increasingly sophisticated tactics while security findings proliferate across multiple channels.
Security personnel struggle to process high volumes of textual reports~\citep{hinchy_voice_2022}, impeding their primary mission of threat identification and infrastructure protection.
Despite existing frameworks like the \ac{cti} cycle~\citep{sauerwein_threat_2021} and automation methods~\citep{kuehn_ovana_2021}, information processing challenges persist.
While \ac{cti} -- the process of collecting and analyzing security data to derive actionable recommendations -- can be aggregated in \acp{tip}~\citep{preuveneers_privacypreserving_2022}, the diversity of sources and evolving threats creates significant information overload~\citep{kaufhold_rapid_2020}.

{
\acp{cert}, as organizational security incident coordinators~\citep{skopik_problem_2016}, require current threat intelligence for effective response.
Studies reveal that \num{45}{\si{\percent}} of \ac{cert} teams process only critical reports due to understaffing~\citep{gorzelak_proactive_2011}, while \num{13}{\si{\percent}} lack capacity for new information and \num{11}{\si{\percent}} cannot manage existing volumes.
Recent research~\citep{hinchy_voice_2022,kaufhold_we_2024} reinforces these challenges, with \num{47.6}~{\si{\percent}} of analysts reporting burnout and \num{46.6}{\si{\percent}} identifying threat monitoring as their most time-consuming task.
For \num{19.2}~{\si{\percent}} of analysts, automating threat alert enrichment through incident correlation represents a critical priority~\citep{hinchy_voice_2022}.
\citet{kaufhold_we_2024} highlight persistent manual processes in technical information exchange, redundancy checks, and general automation needs, underscoring the urgency for enhanced information processing solutions.
\looseness=-1
}

\paragraph{Goal}
This research evaluates clustering algorithms' efficacy in supporting \ac{cert} threat information processing.
Clustering enables efficient threat analysis by allowing rapid overview of related data points before detailed investigation.
We assess various embedding-clustering algorithm combinations against derived requirements, with particular emphasis on threat messages and security advisories from both commercial vendors and security researchers.
This investigation addresses our primary research question: \emph{Which cluster algorithm and embedding combination is suitable to reduce \ac{cert} personnel's information overload}~(\rq{})?

\paragraph{Contributions}
This work advances current research through two primary contributions:
(i)~introduction of \threatreport{}, a novel labeled threat report corpus~(\c{1}) and
(ii)~a comprehensive performance comparisons of 14 clustering algorithms on the created embeddings across the five diverse datasets~(\c{2}).

\paragraph{Outline}
The remainder of this paper is structured as follows:
\cref{clustering:sec:related_work} examines related work and identifies research gaps.
\cref{clustering:sec:method} details our methodology, followed by our comprehensive evaluation results in \cref{clustering:sec:evaluation}.
\cref{clustering:sec:discussion} discusses findings and limitations, while \cref{clustering:sec:conclusion} summarizes our contributions.

\section{Related Work}
\label{clustering:sec:related_work}

{
	We present related work in embeddings, clustering, and evaluation, culminating in the identification of our research gap.
	\looseness=-1
}

\paragraph{Embeddings}
Embedding methods transform data points into vector representations where similarity is preserved through spatial proximity.
These range from simple word frequency approaches to sophisticated language models encoding semantic relationships~\citep{zong_text_2021,reimers_sentencebert_2019}.
Document-level encoding presents unique challenges for threat intelligence processing.
Traditional approaches include \ac{bow}, which records absolute term frequencies using a global vocabulary, and \ac{tfidf}, which weights terms by their document frequency~\citep{zong_text_2021}.
Recent approaches use \acs{bert}~\citep{devlin_bert_2018}, with \ac{sbert} specifically optimized for embedding longer text units~\citep{reimers_sentencebert_2019}.
The \emph{MTEB} benchmark provides comprehensive performance comparisons of differently-sized \acp{llm}, including clustering efficacy~\citep{muennighoff_mteb_2023}.

\paragraph{Clustering}
Clustering algorithms group data points based on similarity metrics such as cosine distance or silhouette scores~\citep{ignaczak_text_2021,zong_text_2021,rousseeuw_silhouettes_1987}.
Traditional methods range from centroid-based K-Means~\citep{lloyd_least_1982} requiring predefined cluster counts to density-based \ac{dbscan}~\citep{ester_densitybased_1996} supporting arbitrary cluster shapes.
Recent research explores deep learning approaches that leverage intermediate representations~\citep{mautz_deep_2019,mcconville_n2d_2021,leiber_dipencoder_2022,leiber_dipbased_2021,ren_deep_2020,moradifard_deep_2020,xie_unsupervised_2016}.
In security contexts, clustering facilitates log summarization~\citep{gove_automatic_2022}, Android permission analysis~\citep{milosevic_machine_2017}, and cybersecurity event detection in social media~\citep{riebe_cysecalert_2021} using various techniques from local sensitivity hashing to neural networks.
Vulnerability management benefits from clustering through alternative vulnerability classification~\citep{anastasiadis_combining_2022}.

\paragraph{Evaluation}
Text clustering evaluation employs both internal metrics~(assessing compactness and separability without ground truth)~\citep{rendon_internal_2011} and external metrics~(requiring labeled data)~\citep{zong_text_2021}.
\citet{rosenberg_vmeasure_2007} highlight limitations of traditional metrics like purity and entropy, particularly for edge cases.
The V-measure framework~\citep{rosenberg_vmeasure_2007} combines homogeneity~(cluster label consistency) and completeness~(label distribution) metrics, providing comprehensive clustering quality assessment.
Recent frameworks~\citep{leiber_benchmarking_2023} integrate multiple algorithms, datasets, and metrics for systematic evaluation.

\paragraph{Research Gap}
While existing research addresses clustering of security information, it primarily focuses on short-form content~(\eg, social media posts)~\citep{riebe_cysecalert_2021} or traditional embedding methods~\citep{riebe_cysecalert_2021,lesceller_sonar_2017}.
No comprehensive evaluation exists, which compares modern embedding-clustering combinations for longer security texts, such as security advisories or threat reports.
This gap is particularly significant given the increasing volume and complexity of security documentation requiring efficient processing by \ac{cert} personnel.

\section{Methodology}
\label{clustering:sec:method}
We present the data used in this work and the requirements for document embeddings, clustering algorithms, and evaluation metrics.

\begin{table}
	\caption{
		This table outlines the structural information of the datasets $c \in [\mathit{CySecAlert}, \mathit{MSE}, \mathit{ThreatReport}, \mathit{SBR}, \mathit{SMS}]$.
		$L_c$ is the sequence~$\mathit{len}(\mathit{dp}_i)$ in character for all $\mathit{dp}_i \in c$.
		It shows the size~$\abs{L_c}$, the average length~\avg{L_c}, the median~\median{L_c}, the minimum~\mymin{L_c} and maximum~\mymax{L_c} data point length, and the number of ground truth clusters~(\emph{\#$L_c$}) of $c$.
	}
	\label{clustering:tab:corpusStructure}

	\sisetup{table-auto-round}
	\begin{tabular}{
			@{}
			r
			S[table-format=5.0]
			S[table-format=4.0]
			S[table-format=4.0]
			S[table-format=2.0]
			S[table-format=5.0]
			S[table-format=2.0]
			@{}
		}
		\toprule
		{$c$}         & {\abs{L_c}}  & {\avg{L_c}} & {\median{L_c}} & {\mymin{L_c}} & {\mymax{L_c}} & {\emph{\#$L_c$}} \\
		\midrule
		\cysecalert   & 13306.000000 & 136.429806  & 119.000000     & 6.000000      & 486.000000    & 2.000000         \\
		MSE           & 3001.000000  & 283.604465  & 277.000000     & 57.000000     & 686.000000    & 2.000000         \\
		\threatreport & 461.000000   & 4369.967462 & 3366.000000    & 7.000000      & 26853.000000  & 39.000000        \\ \addlinespace
		SBR           & 5000.000000  & 887.097200  & 457.500000     & 29.000000     & 32785.000000  & 5.000000         \\ \addlinespace
		SMS           & 5574.000000  & 80.445102   & 61.000000      & 2.000000      & 910.000000    & 2.000000         \\
		\bottomrule
	\end{tabular}
\end{table}

\subsection{Text Corpora}
\label{clustering:sec:textcorpus}

This study employs multiple datasets to evaluate the selected clustering algorithms across three distinct use cases:
(I)~effectiveness in processing threat-related short messages and threat reports,
(II)~performance in handling \ac{sbr} across diverse products, and
(III)~comparative analysis on non-security short messages.
Exemplar texts from each corpus are presented in Listing~\ref{clustering:fig:exampletexts}, while \cref{clustering:tab:corpusStructure} provides a comprehensive overview of the datasets' structural characteristics.

For security-centric analysis, we utilize three primary datasets: \cysecalert{}~\citep{riebe_cysecalert_2021}, \ac{mse}~\citep{bayer_multilevel_2023}, and \threatreport{}~(self-labeled).
The \cysecalert{} and \ac{mse} datasets comprise security-related short messages extracted from X~(formerly Twitter).
The \threatreport{} dataset encompasses security-related content aggregated from news outlets and security feeds.
While the former two datasets are representative for \ac{cert} data aggregations in crisis, the third represents data of the daily work of \acp{cert}.
In both areas the volume of information increased tremendously in recent years, while understaffing remained on a high level~\citep{isaca_companies_2024}.
For product-specific analysis, the \ac{sbr} dataset contains security-related messages from issue trackers spanning five distinct products~\citep{shu_better_2019}.
Both use-cases
To establish a baseline for general text classification, we incorporate the UCI \emph{SMS Spam Collection}~\citep{almeida_sms_2012}, which features characteristics common to security domain texts, including abbreviations, non-standard nomenclature, and spam content.

The labeling of \threatreport{} was done by two researchers in the field of information security.
After the first independent labeling of $10$ data points, both researchers discussed their labeling process and aligned differences.
Afterward, both researchers continued independent on a half dataset each.

\begin{figure}
	\setcaptiontype{lstlisting}

	\begin{subfigure}{\linewidth}
		\centering
		\myquote{CyberRange : The Open-Source AWS Cyber Range [\dots]}
		\caption{Example text for the \cysecalert{} dataset (use-case I).}
		\label{clustering:fig:examplecysecalert}
	\end{subfigure}\vspace{1em}

	\begin{subfigure}{\linewidth}
		\centering
		\myquote{SMBs need to take immediate action on \#microsoft \#exchange \#vulnerabilities [URL] [\dots]}
		\caption{Example text for the \acl{mse} dataset (use-case I).}
		\label{clustering:fig:examplemse}
	\end{subfigure}\vspace{1em}

	\begin{subfigure}{\linewidth}
		\centering
		\myquote{New CacheWarp AMD CPU attack lets hackers gain root in Linux VMs- November 14, 2023- 03:34 PM-2 A new software-based fault injection attack, CacheWarp, can let threat actors hack into AMD SEV-protected [\dots]}
		\caption{Example text for the \threatreport{} dataset (use-case I).}
		\label{clustering:fig:examplereport}
	\end{subfigure}\vspace{1em}

	\begin{subfigure}{\linewidth}
		\centering
		\myquote{\texttt{SYSCS\_UTIL.SYSCS\_COMPRESS\_TABLE} should create statistics if they do not exist There must be an entry in the \texttt{SYSSTATISTICS} table in order for the cardinality statistics in \texttt{SYSSTATISTICS} to be created with \texttt{SYSCS\_UTIL.SYSCS\_COMPRESS\_TABLE SYSCS\_UTIL.SYSCS\_COMPRESS\_TABLE} should create statistics if they don't exist. [\dots]}
		\caption{Example text for the \acl{sbr} dataset (use-case II).}
		\label{clustering:fig:examplesbr}
	\end{subfigure}\vspace{1em}

	\begin{subfigure}{\linewidth}
		\centering
		\myquote{Auction round 4. The highest bid is now £54. Next maximum bid is £71. To bid, send BIDS e. g. 10 (to bid £10) to 83383. Good luck}
		\caption{Example text for the SMS dataset (use-case III).}
		\label{clustering:fig:examplesms}
	\end{subfigure}

	\caption{Example texts from the different evaluation datasets~(\cysecalert{}, MSE, \threatreport, \acs*{sbr}, and SMS).}
	\label{clustering:fig:exampletexts}
\end{figure}

\subsection{Operational Requirements for Automated \acs*{cert} Information Processing}
\label{clustering:subsec:requirements}

{
	The exponential growth in security-related data requires \acp{cert} to conduct extensive manual analysis daily, leading to significant operational strain~\citep{hinchy_voice_2022,kaufhold_we_2024}.
	This sustained cognitive load frequently results in professional burnout and potential workforce attrition.
	The automation of routine analytical tasks presents a critical opportunity for operational improvement, particularly in the identification of duplicate and related information within incoming data streams.
	Research indicates that security personnel estimate \say{half of their tasks to all of their tasks could be automated today}~\citep{hinchy_voice_2022}.
	Drawing from multiple empirical studies~\citep{hinchy_voice_2022,basyurt_help_2022,kaufhold_we_2024}, we establish the following core requirements for an effective \ac{cert} clustering system.
	\looseness=-1
}

\begin{enumerate}[label={\textbf{\emph{R\arabic*}}}]
	\item \textbf{Reducing information overload for \acp{cert}.}:
	      The clustering system must demonstrably reduce the volume of information requiring manual review through effective cluster consolidation.
	      Cluster homogeneity must be maximized through rigorous outlier management.
	      The presence of misclassified data points would significantly compromise cluster integrity and negate the intended benefits of information reduction.
	      Therefore, the system must prioritize classification accuracy over cluster completeness.
	\item \textbf{Unburden \acp{cert}.}:
	      The proposed algorithms must operate with minimal configuration requirements, eliminating the need for continuous model adjustment with emerging vulnerabilities or technologies.
	      Model fine-tuning represents a significant operational overhead.
	      While potentially more engaging than routine document review, such tasks divert resources from core responsibilities: threat identification, analysis, and stakeholder communication.
	\item \textbf{Retention of data.}:
	      All alerts must remain accessible, regardless of their cluster assignability.
	      The system must preserve outliers during analysis rather than forcing them into inappropriate clusters.
	      Otherwise, important information might be either missed, due to being assigned to an outlier cluster or simply confuses personnel if found in wrong clusters.
	      Either case would diminish the benefits of clustering the data due to lost trust in the system.
	      This requirement aligns with information overload reduction by enabling a discrete outlier cluster for manual review, rather than discarding or misclassifying these data points.
	\item \textbf{Runtime performance.}:
	      While not the primary optimization target, the system must complete clustering operations on new data and present results within an operationally acceptable timeframe, defined here as several minutes.
	      The system might be run on-demand by \ac{cert} personnel in preparation of the daily inbound information review.
\end{enumerate}

\subsection{Embeddings, Clustering, and Evaluation}
\label{clustering:subsec:embeddings_clustering_evaluation}

\begin{table}
	\def\hfurl#1{\href{https://huggingface.co/#1}{#1}}
	\caption{Overview of the \acp{llm} used in combination with SBERT~(sorted alphabetically).}
	\label{clustering:tab:models}
	\begin{tabular}{ll}
		\toprule
		\textbf{Huggingface Model ID~$\downarrow$}      & \textbf{Params} \\
		\midrule
		\hfurl{Alibaba-NLP/gte-base-en-v1.5}            & 137M            \\
		\hfurl{jxm/cde-small-v1}                        & 281M            \\
		\hfurl{markusbayer/cysecbert}                   & 110M            \\
		\hfurl{meta-llama/Llama-3.2-1B}                 & 1.24B           \\
		\hfurl{meta-llama/Llama-3.2-3B}                 & 3.21B           \\
		\hfurl{mistralai/Mistral-7B-v0.1}               & 7.24B           \\
		\hfurl{mistralai/Mistral-7B-v0.3}               & 7.25B           \\
		\hfurl{NovaSearch/stella\_en\_1.5b\_v5}         & 1.54B           \\
		\hfurl{NovaSearch/stella\_en\_400M\_v5}         & 435M            \\
		\hfurl{nvidia/NV-Embed-v2}                      & 7.85B           \\
		\hfurl{sentence-transformers/all-MiniLM-L12-v1} & 33.4M           \\
		\hfurl{sentence-transformers/all-mpnet-base-v2} & 109M            \\
		\hfurl{thenlper/gte-large}                      & 335M            \\
		\hfurl{sentence-transformers/gtr-t5-xxl}        & 4.86B           \\
		\bottomrule
	\end{tabular}
\end{table}

\begin{table*}
	\caption{%
		Overview of used cluster algorithms, a short explanation, and the used parameters.
		If no parameter setting is displayed, we use the defaults selected by ClustPy~\citep{leiber_benchmarking_2023}.
	}
	\label{clustering:tab:clusteralgos}

	\centering
	{
		\begin{tabular}{@{} r p{.55\linewidth} l @{}}
			\toprule
			\textbf{Algorithm}                                        & \textbf{Description}                                                                                                                                                              & \textbf{Parameters}            \\
			\midrule
			\multicolumn{3}{c}{\emph{Partition-based Clustering}}                                                                                                                                                                                                                          \\
			\midrule
			K-Means~\citep{zong_text_2021}                            & Divides data into $k$ clusters by minimizing within-cluster variance. Uses centroid-based approach with spherical cluster assumptions.                                            & n\_cluster = 12                \\
			Bruteforce-K-Means                                        & Bruteforce to find optimal $k \in \left[ 2, \mymax{\sqrt{\abs{c}}} \right]$ based on resulting silhouette score, where $c$ is the given dataset.                                                                   \\
			SpecialK~\citep{hess_magic_2020}                          & Determining the optimal number of clusters by developing a probabilistic method to assess if clusters originate from a single distribution                                                                         \\
			SkinnyDip~\citep{maurus_skinnydip_2016}                   & Noise-robust clustering algorithm designed for datasets with up to 80~\% noise, using Hartigan's dip test of unimodality and recursive univariate projection analysis.                                             \\
			\midrule
			\multicolumn{3}{c}{\emph{Hierarchical Clustering}}                                                                                                                                                                                                                             \\
			\midrule
			AgglomerativeClustering~\citep{nielsen_hierarchical_2016} & Hierarchical bottom-up clustering merging closest data points, creating cluster hierarchy via dendrogram.                                                                         & n\_clusters = 12               \\
			\midrule
			\multicolumn{3}{c}{\emph{Density-based Clustering}}                                                                                                                                                                                                                            \\
			\midrule
			DBSCAN~\citep{ester_densitybased_1996}                    & Clusters dense regions separated by low-density areas. Robust to outliers, discovers arbitrarily shaped clusters                                                                  & \makecell[tl]{eps = 0.5        \\ min\_samples = 2\\ metric = precomputed} \\
			OPTICS~\citep{ankerst_optics_1999}                        & Advanced density-based clustering handling varying cluster densities. Creates reachability plot for comprehensive structure analysis.                                             & \makecell[tl]{min\_samples = 5 \\ metric = precomputed} \\
			\midrule
			\multicolumn{3}{c}{\emph{Deep Clustering}}                                                                                                                                                                                                                                     \\
			\midrule
			DKM~\citep{moradifard_deep_2020}                          & Deep clustering approach that jointly learns data representations and cluster assignments through a continuous reparametrization for k-Means, solely relying on gradient descent. & n\_cluster = 12                \\
			DDC~\citep{ren_deep_2020}                                 & Two-stage deep density-based image clustering framework using convolutional autoencoder and t-SNE for low dimensionality and density-based clustering to recognize clusters.                                       \\
			DipEncoder~\citep{leiber_dipencoder_2022}                 & Coupling the Hartigan's unsupervised Dip-test with an autoencoder to obtain cluster embeddings.                                                                                   & n\_cluster = 12                \\
			N2D~\citep{mcconville_n2d_2021}                           & Learns autoencoded embedding, uses UMAP for manifold learning, and applies shallow clustering algorithms.                                                                         & n\_cluster = 12                \\
			DeepECT~\citep{mautz_deep_2019}                           & Builds cluster tree in embedding space, which allows selecting the number of clusters afterwards.                                                                                                                  \\
			DEC~\citep{xie_unsupervised_2016}                         & Neural network clustering with iterative representation and cluster refinement.                                                                                                   & n\_cluster = 12                \\
			DipDECK~\citep{leiber_dipbased_2021}                      & Advanced deep clustering with disentangled representation approach.                                                                                                                                                \\
			\bottomrule
		\end{tabular}
	}
\end{table*}

For the embedding process, we evaluate a diverse range of locally deployable \acp{llm}.
The selected models span from lightweight architectures with \num{33.4}M parameters~(all-MiniLM-L12-v1) to large-scale models with \num{7.85}B parameters~(NV-Embed-v2)~\citep{bayer_cysecbert_2024,reimers_sentencebert_2019,li_general_2023,morris_contextual_2024,liu_spinquant_2024,jiang_mistral_2023,zhang_jasper_2025,moreira_nvretriever_2024,zhang_mgte_2024}.
We perform all embeddings using \emph{sentence-transformers}~\citep{reimers_sentencebert_2019} with the models enumerated in \cref{clustering:tab:models}, utilizing computing infrastructure equipped with either NVidia~A100 or NVidia~H100 GPUs.
Initial attempts to process the embeddings on an Apple~M4 system with \num{24}~GB memory proved insufficient due to memory constraints.

Our clustering methodology incorporates algorithms from four major categories: \emph{partition-based clustering}, \emph{density-based clustering}, \emph{hierarchical clustering}, and \emph{deep clustering}~\citep{zong_text_2021}, selected based on their prevalence, operational characteristics~(\eg, arbitrary cluster shape detection, parameter complexity), and MTEB ranking\footnote{\url{https://huggingface.co/spaces/mteb/leaderboard}}.
We prioritize algorithmic simplicity to assess performance under minimal parameter optimization.
This approach aligns with \r{2}, enabling \acp{cert} to focus on core responsibilities without extensive hyperparameter tuning.
\cref{clustering:tab:clusteralgos} provides a comprehensive overview of the selected clustering algorithms.
To evaluate the requirements~\r{2} and~\r{4} regarding operational burden and runtime performance, we conduct clustering experiments on an Apple M4 system with \num{24}~GB memory.
This represents a worst-case scenario using consumer hardware, below typical \ac{cert} infrastructure capabilities.

The evaluation framework employs exclusively external metrics for two fundamental reasons.
First, internal metrics introduce inherent bias when comparing diverse clustering algorithms, as certain algorithms optimize specific internal criteria~(\eg, K-Means for silhouette coefficient).
Second, intrinsic evaluation metrics assess cluster morphology rather than semantic accuracy relative to ground truth.
For \ac{cert} applications, semantic cohesion within clusters is paramount to prevent analytical confusion and redundant verification.
We adopt the external metrics proposed by~\citet{rosenberg_vmeasure_2007}: \emph{homogeneity}, \emph{completeness}, and \emph{V-measure}, with primary emphasis on \emph{homogeneity}.
Homogeneity~$h \in \left[0, 1 \right] \subseteq \mathbb{R}$ quantifies intra-cluster uniformity, achieving its maximum of \num{1} when clusters perfectly align with ground truth~\citep{rosenberg_vmeasure_2007}.
While completeness~$c \in \left[0, 1 \right] \subseteq \mathbb{R}$ measures object distribution across clusters, our focus on precision renders this metric secondary.
V-measure~$V_\beta = \frac{(1+\beta)hc}{(\beta*h)+c} \subseteq \mathbb{R}$ combines these metrics through their harmonic mean, reaching \num{1} for optimal clustering.
We set $\beta = 0$, effectively reducing V-measure to homogeneity, prioritizing semantic consistency for \ac{cert} operations.

Our comprehensive evaluation encompasses \num{14} embedding models, \num{14} clustering algorithms across \num{5} datasets, measuring \num{8} distinct metrics while prioritizing \emph{homogeneity}.
Additional captured metrics include:
\emph{completeness},
\emph{V-measure},
\emph{silhouette coefficient},
\emph{Adjusted Rand Index},
\emph{Calinski-Harabasz index},
\emph{Davies-Bouldin index}, and
\emph{runtime performance}.
All metrics represent averages across \num{5} consecutive iterations, totaling \num{4900} distinct experimental configurations.

\section{Evaluation}
\label{clustering:sec:evaluation}

{
	Our evaluation methodology comprises three primary components.
	First, we analyze clustering performance across three distinct dataset categories: (i)~\ac{cti} datasets, (ii)~\acp{sbr}, and (iii)~general short message data~(UCI's SMS dataset).
	Second, we assess computational efficiency through runtime analysis of both embedding generation and clustering operations.
	Finally, we correlate these findings with our research questions and established requirements in the subsequent section.
	\looseness=-1
}

\subsection{Clustering Performance}
\label{clustering:subsec:clusteringperformance}

\begin{table*}
	\centering
	\caption{
	Evaluation of the datasets sorted by homogeneity~(H) with ranking~(\#).
	It shows the used evaluation combination: clustering algorithm~(Algorithm), embedding model~(Model), and metrics.
	The columns denote homogeneity~(H), completeness~(C), V-measure~(V-M), Silhouette coefficient~(Sil), Adjusted Rand Index~(ARI), Calinski-Harabasz index~(CH), Davies-Bouldin index~(DB), runtime in seconds~(t~[s]), and number of predicted clusters~(\#C), respectively.
	Results are sorted in descending order by homogeneity~($\downarrow$).
	After the evaluation we evaluated the \threatreport{} dataset with the parameters \emph{n\_clusters = 60}, denoted as \threatreport-60.
	}
	\label{clustering:tab:evaluation}
	\sisetup{table-auto-round}
	\begin{tabular}{lllrrrrrrrrr}
		\toprule
		  & \textbf{Algorithm}      & \textbf{Model}       & \textbf{H}~$\downarrow$ & \textbf{C} & \textbf{V-M} & \textbf{Sil} & \textbf{ARI} & \textbf{CH} & \textbf{DB} & \textbf{t~[s]} & \textbf{\#C} \\

		\midrule
		\multicolumn{12}{c}{\emph{\cysecalert} (GT: \abs{c} = \num{13306}, \#C = 2) --- Use-case I}                                                                                                        \\
		\midrule
		1 & K-Means                 & NV-Embed-v2          & 0.61                    & 0.10       & 0.17         & 0.01         & 0.04         & 177.35      & 5.27        & 2.75           & 12.0         \\
		2 & BruteForceK-Means       & NV-Embed-v2          & 0.59                    & 0.11       & 0.19         & 0.01         & 0.04         & 240.24      & 5.42        & 1.65           & 8.0          \\
		3 & K-Means                 & stella\_en\_400M\_v5 & 0.55                    & 0.09       & 0.15         & 0.02         & 0.04         & 205.96      & 4.82        & 0.84           & 12.0         \\
		4 & K-Means                 & stella\_en\_1.5b\_v5 & 0.54                    & 0.09       & 0.15         & -0.01        & 0.03         & 212.45      & 4.85        & 0.88           & 12.0         \\
		5 & BruteForceK-Means       & stella\_en\_400M\_v5 & 0.54                    & 0.10       & 0.17         & 0.02         & 0.05         & 280.10      & 4.99        & 0.61           & 8.0          \\ \addlinespace

		\midrule
		\multicolumn{12}{c}{\emph{MSE} (GT: \abs{c} = \num{3001}, \#C = 2) --- Use-case I}                                                                                                                 \\
		\midrule
		1 & OPTICS                  & llama-3.2-1b         & 0.49                    & 0.10       & 0.17         & 0.26         & 0.03         & 16.87       & 1.43        & 3.99           & 156.0        \\
		2 & OPTICS                  & all-minilm-l12-v2    & 0.48                    & 0.10       & 0.17         & 0.24         & 0.03         & 17.20       & 1.41        & 4.04           & 150.0        \\
		3 & OPTICS                  & all-mpnet-base-v2    & 0.45                    & 0.10       & 0.16         & 0.21         & 0.03         & 16.81       & 1.40        & 3.99           & 145.0        \\
		4 & OPTICS                  & NV-Embed-v2          & 0.45                    & 0.10       & 0.16         & 0.17         & 0.02         & 13.74       & 1.53        & 4.01           & 161.0        \\
		5 & OPTICS                  & gte-base-en-v15      & 0.45                    & 0.11       & 0.17         & 0.20         & 0.03         & 16.53       & 1.39        & 3.96           & 135.0        \\ \addlinespace

		\midrule
		\multicolumn{12}{c}{\emph{\threatreport} (GT: \abs{c} = \num{461}, \#C = 39) --- Use-case I}                                                                                                       \\
		\midrule
		1 & DDC                     & gte-base-en-v15      & 0.34                    & 0.30       & 0.32         & 0.02         & 0.03         & 6.90        & 2.26        & 2.90           & 30.4         \\
		2 & DDC                     & all-minilm-l12-v2    & 0.33                    & 0.29       & 0.31         & -0.01        & 0.02         & 6.88        & 2.08        & 2.54           & 34.0         \\
		3 & OPTICS                  & gte-large            & 0.32                    & 0.34       & 0.33         & 0.08         & 0.05         & 9.28        & 2.16        & 0.10           & 23.0         \\
		4 & OPTICS                  & stella\_en\_1.5b\_v5 & 0.31                    & 0.36       & 0.34         & 0.01         & 0.06         & 7.67        & 2.01        & 0.14           & 25.0         \\
		5 & OPTICS                  & gte-base-en-v15      & 0.31                    & 0.35       & 0.33         & 0.08         & 0.06         & 7.50        & 2.07        & 0.12           & 25.0         \\ \addlinespace

		\midrule
		\multicolumn{12}{c}{\emph{\threatreport-60} (GT: \abs{c} = \num{461}, \#C = 39) --- Use-case I}                                                                                                    \\
		\midrule
		1 & AgglomerativeClustering & gte-base-en-v15      & 0.51                    & 0.33       & 0.40         & 0.27         & 0.02         & 9.26        & 1.65        & 0.03           & 60           \\
		2 & AgglomerativeClustering & all-mpnet-base-v2    & 0.50                    & 0.33       & 0.40         & 0.27         & 0.02         & 10.59       & 1.55        & 0.03           & 60           \\
		3 & AgglomerativeClustering & stella\_en\_400M\_v5 & 0.50                    & 0.33       & 0.40         & 0.28         & 0.02         & 10.91       & 1.46        & 0.04           & 60           \\
		4 & N2D                     & gte-base-en-v15      & 0.50                    & 0.32       & 0.39         & 0.20         & 0.01         & 7.38        & 1.99        & 2.92           & 60           \\
		5 & AgglomerativeClustering & gtr-t5-xxl           & 0.50                    & 0.32       & 0.39         & 0.26         & 0.02         & 9.25        & 1.63        & 0.02           & 60           \\ \addlinespace

		\midrule
		\multicolumn{12}{c}{\emph{\Acl*{sbr}} (GT: \abs{c} = \num{5000}, \#C = 5) --- Use-case II}                                                                                                         \\
		\midrule
		1 & DEC                     & stella\_en\_1.5b\_v5 & 0.88                    & 0.63       & 0.74         & 0.02         & 0.64         & 91.20       & 4.94        & 39.66          & 12.0         \\
		2 & DKM                     & stella\_en\_1.5b\_v5 & 0.87                    & 0.62       & 0.72         & -0.07        & 0.61         & 87.08       & 5.59        & 33.41          & 12.0         \\
		3 & DeepECT                 & stella\_en\_1.5b\_v5 & 0.86                    & 0.47       & 0.61         & -0.00        & 0.35         & 62.77       & 5.06        & 68.39          & 20.0         \\
		4 & DEC                     & stella\_en\_400M\_v5 & 0.86                    & 0.60       & 0.71         & 0.02         & 0.60         & 96.50       & 6.44        & 39.52          & 12.0         \\
		5 & DEC                     & gte-base-en-v15      & 0.84                    & 0.58       & 0.68         & 0.04         & 0.57         & 93.44       & 4.97        & 34.88          & 12.0         \\ \addlinespace

		\midrule
		\multicolumn{12}{c}{\emph{SMS} (GT: \abs{c} = \num{5574}, \#C = 2) --- Use-case III}                                                                                                               \\
		\midrule
		1 & K-Means                 & NV-Embed-v2          & 0.89                    & 0.15       & 0.25         & 0.02         & 0.06         & 78.78       & 4.73        & 2.42           & 12.0         \\
		2 & DKM                     & NV-Embed-v2          & 0.88                    & 0.46       & 0.60         & 0.01         & 0.76         & 38.83       & 5.90        & 64.98          & 12.0         \\
		3 & BruteForceK-Means       & NV-Embed-v2          & 0.88                    & 0.18       & 0.29         & 0.03         & 0.09         & 103.05      & 4.95        & 1.22           & 8.0          \\
		4 & DDC                     & NV-Embed-v2          & 0.88                    & 0.24       & 0.37         & 0.01         & 0.27         & 77.20       & 4.76        & 53.21          & 8.8          \\
		5 & DipEncoder              & NV-Embed-v2          & 0.88                    & 0.14       & 0.24         & 0.02         & 0.05         & 71.92       & 5.44        & 159.85         & 12.0         \\
		\bottomrule
	\end{tabular}
\end{table*}

Our analysis encompasses \num{196} distinct cluster-embedding combinations for each dataset, with results averaged across \num{5} consecutive executions.
Clustering performance exhibits significant variation correlated with ground truth cluster cardinality and dataset dimensionality.
\cref{clustering:tab:evaluation} presents the five highest-performing cluster-embedding combinations per dataset.

The \cysecalert{} dataset demonstrates superior performance with partition-based clustering methodologies.
Model capacity shows limited correlation with performance, as evidenced by comparable results between \emph{stella\_en\_400M\_v5}~(\num{435}M parameters) and \emph{NV-Embed-v2}~(\num{7.85}B parameters).
While homogeneity metrics achieve a maximum of \num{.61}, additional metrics indicate significant cluster overlap.
Completeness remains below \num{.11}, yielding a maximum V-measure of \num{.19}.
Silhouette coefficient, Adjusted Rand Index, and Calinski-Harabasz index collectively indicate suboptimal cluster separation.
The optimal configuration generates \num{12} clusters, achieving a \num{99.91}~\% reduction in dataset cardinality.

For the MSE dataset, density-based clustering, specifically OPTICS, achieves optimal performance.
Homogeneity reaches \num{.49}, with superior performance from models below \num{1.24}B parameters (Llama-3.2-1B) to \num{33}M parameters.
Auxiliary metrics suggest cluster overlap challenges, while achieving \num{94.80}~\% input reduction.

The \threatreport{} dataset exhibits optimal performance with deep and density-based clustering, achieving maximum homogeneity of \num{.34}.
These results indicate insufficient performance for production deployment, necessitating enhanced fine-tuning procedures.
Supplementary metrics corroborate suboptimal clustering performance.
Despite achieving \num{93.41}~\% dimensional reduction, the clustering quality remains inadequate for operational deployment.

Deep clustering demonstrates superior performance on the \ac{sbr} dataset, with \emph{stella} models~\see{clustering:tab:models} achieving homogeneity exceeding \num{.86}.
This performance suggests that large-scale models like \emph{NV-Embed-v2} are not prerequisite for optimal embedding generation.
The configuration achieves \num{99.76}~\% dimensional reduction while maintaining high homogeneity.
However, auxiliary metrics indicate persistent cluster overlap challenges, with silhouette coefficients approximating \num{0}.

\begin{figure*}
	\begin{subfigure}{.62\linewidth}
		\includegraphics[width=\linewidth]{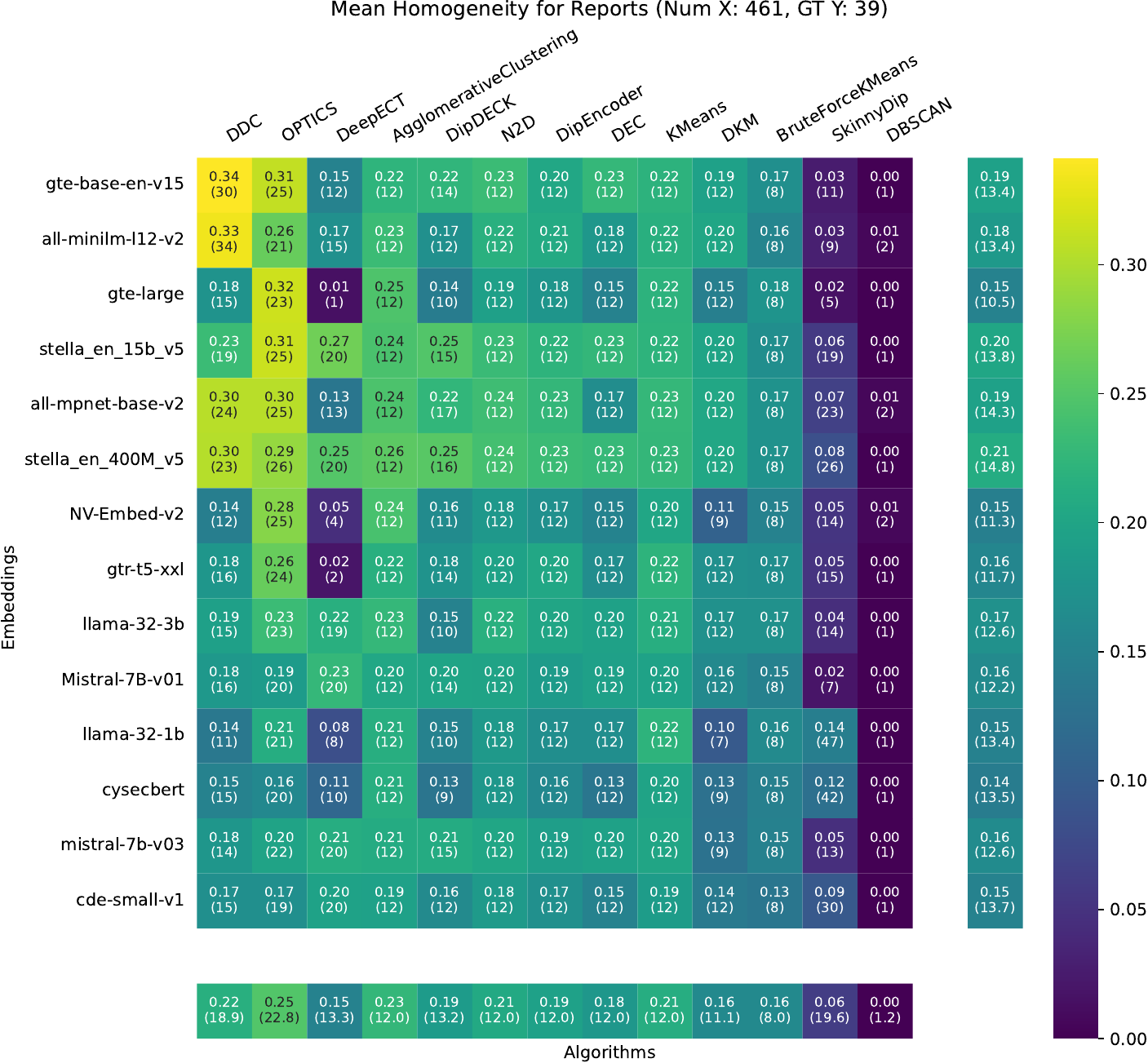}
		\caption{Mean homogeneity of the \threatreport{} dataset.}
		\label{clustering:fig:threatreport_homogeneity}
	\end{subfigure}
	\vspace{2em}

	\begin{subfigure}{.62\linewidth}
		\includegraphics[width=\linewidth]{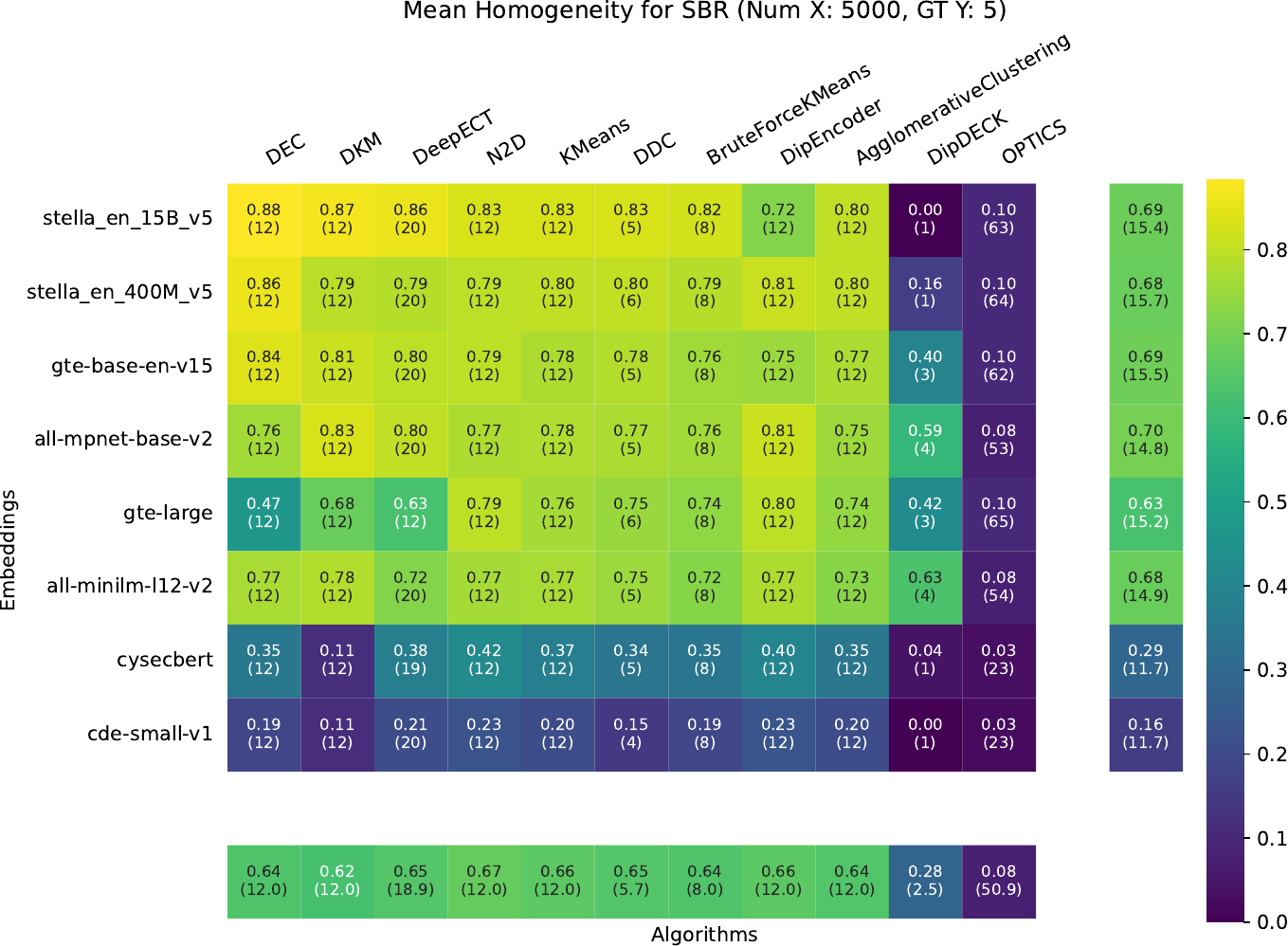}
		\caption{Mean homogeneity of the \ac{sbr} dataset.}
		\label{clustering:fig:sbr_homogeneity}
	\end{subfigure}
	\caption{
		Result of the best and worst performing security related datasets with regard to the mean homogeneity over \num{5} consecutive runs.
		Columns show the clustering algorithms and rows the used embeddings.
		The separated column and row depict the mean over each column and row, respectively.
		The models and algorithms are sorted by the rows/columns sum~(descending), such that the top left shows the highest result, while the bottom right shows the worst.
	}
	\label{clustering:fig:homogeneity}
\end{figure*}

For the general-purpose SMS dataset, partition-based and deep clustering algorithms paired with \emph{NV-Embed-v2} achieve optimal performance.
K-Means clustering achieves \num{99.78}~\% input reduction with exceptional homogeneity exceeding \num{.88} across top-five configurations.
However, secondary metrics continue to indicate cluster overlap challenges.
\cref{clustering:fig:homogeneity} depicts the best and worst performing security related datasets.
The rest is depicted in the \nameref{sec:appendix}.

\begin{table*}
	\caption{%
		Runtime results using the mean value of a total of 5 iterations for all datasets.
	}
	\label{clustering:tab:runtime-reports}

	\begin{tabular}{lrrrrrrrrrrrrrr}
		     & \rotatebox{90}{K-Means} & \rotatebox{90}{AgglomerativeClustering} & \rotatebox{90}{OPTICS} & \rotatebox{90}{BruteForceK-Means} & \rotatebox{90}{SkinnyDip} & \rotatebox{90}{DDC} & \rotatebox{90}{DEC} & \rotatebox{90}{DeepECT} & \rotatebox{90}{DipDECK} & \rotatebox{90}{DipEncoder} & \rotatebox{90}{DKM} & \rotatebox{90}{N2D} & \rotatebox{90}{DBSCAN} & \rotatebox{90}{SpecialK} \\

		\midrule
		\multicolumn{15}{c}{\emph{\cysecalert} (GT: \abs{c} = \num{13306}, \#C = 2) --- Use-case I}                                                                                                                                                                                                                                                                                                  \\
		\midrule
		Min  & 0.54                   & 8.08                                    & 302.67                 & 0.40                             & 0.06                      & 76.83               & 92.82               & 180.59                  & 56.18                   & 219.19                     & 78.87               & 73.20               & -                      & -                        \\
		Max  & 4.30                   & 80.28                                   & 313.88                 & 2.94                             & 0.40                      & 131.88              & 184.92              & 377.33                  & 186.90                  & 344.12                     & 174.94              & 133.67              & -                      & -                        \\
		Mean & 1.51                   & 37.04                                   & 307.16                 & 1.09                             & 0.18                      & 95.09               & 123.81              & 236.20                  & 113.20                  & 265.79                     & 109.43              & 92.34               & -                      & -                        \\

		\midrule
		\multicolumn{15}{c}{\emph{MSE} (GT: \abs{c} = \num{3001}, \#C = 2) --- Use-case I}                                                                                                                                                                                                                                                                                                           \\
		\midrule
		Min  & 0.11                   & 0.45                                    & 3.96                   & 0.07                             & 1.05                      & 14.40               & 17.65               & 38.47                   & 27.49                   & 55.66                      & 15.43               & 15.07               & 0.07                   & -                        \\
		Max  & 1.26                   & 4.02                                    & 4.25                   & 1.05                             & 13.16                     & 28.29               & 41.53               & 111.69                  & 58.35                   & 85.04                      & 37.79               & 28.87               & 0.10                   & -                        \\
		Mean & 0.45                   & 1.80                                    & 4.03                   & 0.32                             & 5.42                      & 19.39               & 26.34               & 55.90                   & 35.69                   & 67.15                      & 23.04               & 19.63               & 0.08                   & -                        \\

		\midrule
		\multicolumn{15}{c}{\emph{\threatreport} (GT: \abs{c} = \num{461}, \#C = 39) --- Use-case I}                                                                                                                                                                                                                                                                                                 \\
		\midrule
		Min  & 0.04                   & 0.01                                    & 0.10                   & 0.04                             & 0.09                      & 2.54                & 3.42                & 3.68                    & 6.90                    & 6.02                       & 2.53                & 2.73                & 0.00                   & -                        \\
		Max  & 0.32                   & 0.15                                    & 0.20                   & 0.21                             & 1.31                      & 4.78                & 8.15                & 15.24                   & 14.11                   & 12.07                      & 6.50                & 5.94                & 0.00                   & -                        \\
		Mean & 0.11                   & 0.06                                    & 0.13                   & 0.08                             & 0.48                      & 3.37                & 4.88                & 8.57                    & 9.25                    & 8.38                       & 4.02                & 3.59                & 0.00                   & -                        \\

		\midrule
		\multicolumn{15}{c}{\emph{\Acl*{sbr}} (GT: \abs{c} = \num{5000}, \#C = 5) --- Use-case II}                                                                                                                                                                                                                                                                                                   \\
		\midrule
		Min  & 0.15                   & 1.13                                    & 16.38                  & 0.12                             & -                         & 25.29               & 31.93               & 61.09                   & 21.89                   & 91.79                      & 27.46               & 25.26               & -                      & -                        \\
		Max  & 0.74                   & 2.99                                    & 17.33                  & 1.17                             & -                         & 30.70               & 39.66               & 88.85                   & 48.51                   & 105.60                     & 33.83               & 29.55               & -                      & -                        \\
		Mean & 0.37                   & 2.36                                    & 16.99                  & 0.34                             & -                         & 28.24               & 36.80               & 70.54                   & 36.37                   & 95.85                      & 31.33               & 28.09               & -                      & -                        \\

		\midrule
		\multicolumn{15}{c}{\emph{SMS} (GT: \abs{c} = \num{5574}, \#C = 2) --- Use-case III}                                                                                                                                                                                                                                                                                                         \\
		\midrule
		Min  & 0.25                   & 1.38                                    & 140.46                 & 0.16                             & 0.03                      & 29.14               & 35.31               & 76.86                   & 42.70                   & 107.83                     & 29.81               & 27.18               & 1.28                   & 4.20                     \\
		Max  & 2.49                   & 13.43                                   & 176.86                 & 1.59                             & 1.01                      & 55.41               & 78.64               & 176.91                  & 94.06                   & 171.19                     & 73.11               & 58.07               & 1.44                   & 21.56                    \\
		Mean & 0.92                   & 6.70                                    & 149.56                 & 0.69                             & 0.34                      & 39.96               & 54.16               & 112.33                  & 64.32                   & 128.59                     & 46.51               & 40.02               & 1.39                   & 13.83                    \\
		\bottomrule
	\end{tabular}
\end{table*}

\subsection{Runtime Performance}
\label{clustering:subsec:runtime}

Runtime analysis encompasses both embedding generation and clustering operations, measured in seconds.
Both phases demonstrate acceptable computational efficiency~\see{clustering:tab:evaluation,clustering:tab:runtime-embedding}.
Embedding generation peaks at \num{287}~s for the \cysecalert{} dataset, while requiring only \num{80}~s for the SMS dataset.
The \cysecalert{} dataset exhibits minimum embedding time of \num{14.31}~s with mean execution time of \num{84.15}~s.
When combined with clustering operations, DipEncoder requires maximum \num{344.12}~s, while K-Means achieves optimal performance in \num{4.3}~s, yielding total pipeline execution under \num{300}~s.
Comparable performance characteristics are observed across optimal cluster-embedding combinations:
MSE achieves maximal runtime of \num{235}~s~(embed $+$ OPTICS),
\threatreport{} requires \num{220}~s~(embed $+$ DDC),
\ac{sbr} completes in \num{300}~s~(embed $+$ DeepECT),
and SMS processing concludes in \num{155}~s~(embed $+$ DKM).

\begin{table}
	\caption{Tabular display of runtime statistics of embedding the different datasets.}
	\label{clustering:tab:runtime-embedding}

	\centering
	\begin{tabular}{rrrrrr}
		\toprule
		\textbf{Dataset} & \textbf{Min} & \textbf{Max} & \textbf{Mean} & \textbf{Median} & \textbf{Std} \\
		\midrule
		\cysecalert      & 14.31        & 287.40       & 84.15         & 48.58           & 93.60        \\
		MSE              & 5.01         & 230.32       & 51.72         & 17.86           & 68.36        \\
		\threatreport    & 4.05         & 215.21       & 55.26         & 24.90           & 75.88        \\
		SBR              & 11.18        & 210.71       & 68.47         & 40.83           & 63.69        \\
		SMS              & 7.28         & 80.53        & 32.89         & 23.86           & 27.44        \\
		\bottomrule
	\end{tabular}
\end{table}

\section{Discussion, Limitations, and Future Work}
\label{clustering:sec:discussion}

This research evaluates the efficacy of clustering methods in mitigating information overload for \ac{cert} personnel.
While our evaluation demonstrates promising results, several aspects warrant detailed discussion and highlight opportunities for future research.

\subsection{Discussion}

Our research question \say{Which cluster algorithm and embedding combination is suitable to reduce CERT personnel's information overload?} can be answered with qualified success.
The evaluation demonstrates that no single combination universally excels across all \ac{cti} datasets, though several configurations show promising results.
For immediate operational deployment, deep clustering combined with \emph{stella} models achieves optimal performance on \ac{sbr} data~(homogeneity \num{.88}), while partition-based clustering with \emph{NV-Embed-v2} performs well on general security data~(\cysecalert{}, homogeneity \num{.61}).
However, the core \ac{cert} focus on threat reports~(\threatreport{}) requires careful parameter tuning, achieving homogeneity of \num{.5} only with adjusted cluster counts.

Information overload represents a primary factor in \ac{cert} personnel burnout and attrition~\citep{hinchy_voice_2022}.
To address this challenge, we established clustering requirements through comprehensive analysis of current research~\citep{kaufhold_we_2024,basyurt_help_2022,hinchy_voice_2022,gorzelak_proactive_2011}.
Our methodology incorporated both established and novel clustering approaches, leveraging diverse \acp{llm} for document embedding~\see{clustering:tab:models,clustering:tab:clusteralgos}.
The evaluation framework prioritized homogeneity as an external metric~\citep{rosenberg_vmeasure_2007}, analyzing three distinct \ac{cti} datasets~\citep{riebe_cysecalert_2021,bayer_multilevel_2023}, including our novel \threatreport{} dataset~(\c{1}), alongside \ac{sbr} data~\citep{wu_data_2022} and the UCI SMS corpus~\citep{almeida_sms_2012}.

Our results indicate that achieving homogeneous clustering for \ac{cti} advisories and reports presents significant challenges.
While \cysecalert{} and MSE datasets achieve homogeneity scores of \num{.61} and \num{.49} respectively, these metrics do not fully address primary \ac{cert} operational requirements.
The \threatreport{} dataset achieves only \num{.34} homogeneity with default parameters~\see{clustering:tab:clusteralgos}, resulting in mixed clusters and complex structures~(indicated by near-zero silhouette scores).
A subsequent analysis with adjusted parameters~(n\_clusters $= 60$ versus default \num{12}) for the \threatreport{} dataset yields substantially improved results, surpassing \num{.5} homogeneity and approaching \cysecalert{} and MSE performance~(\cf~\threatreport-60 in \cref{clustering:tab:evaluation}).
This indicates, that the clustering algorithms require further parameter optimizations for the \threatreport{} dataset, beyond simple adjustments like setting \emph{n\_clusters}.
The \ac{sbr} dataset demonstrates exceptional potential with homogeneity exceeding \num{.88}, though cluster structure remains complex~(silhouette scores approximating \num{0}).
Deep clustering methodologies effectively address these structural challenges, producing promising results.
These findings suggest clustering offers domain-specific utility~(\eg, \acl{sbr}), while \ac{cert} applications may require algorithmic fine-tuning or careful dataset curation.
The methodology demonstrates strong generalization, evidenced by SMS dataset homogeneity nearing \num{.9}.

All evaluated algorithms achieve data reduction exceeding \num{90}~\%, directly addressing requirement~\r{1} for information overload mitigation.
Even with adjusted parameters, the \threatreport{} dataset maintains reduction rates above \num{86.98}~\%, validating clustering efficacy in \ac{cert} operations.
The simplified configuration aligns with requirement~\r{2}, reducing operational burden while preserving original data points~\r{3}.

Runtime performance exceeds expectations, with complete processing requiring approximately \num{5} minutes.
This efficiency could potentially save a single \ac{cert} over \num{3750} hours annually\footnote{
	Respondents stated a daily workload of up to $2$ hours~\citep{kaufhold_we_2024}, which is reduced to $5$ minutes and can be run asynchronously plus at most $30$ minutes information screening: $250 \times 1.5 \text{ hours} = 3750 \text{ hours}$.
}.
This performance satisfies requirement~\r{4} while preserving operational capacity for core \ac{cert} infrastructure security responsibilities.

\subsection{Limitations and Future Work}

Our methodology exhibits several key limitations.
Parameter optimization was intentionally omitted to align with requirement~\r{2}, though results suggest its necessity for heterogeneous \ac{cti} report processing.
Our evaluation methodology is constrained by the limited number of consecutive runs~(\num{5}), hardware restrictions~(Apple M4, \num{24}~GB memory), and consistently poor cluster separation indicated by silhouette scores approximating \num{0}.

Dataset limitations include potential bias in the manually labeled \threatreport{} dataset and significant performance sensitivity to cluster count parameters~(\num{12} versus \num{60} clusters).
While achieving high dimensional reduction~(\num{>90}~\%), the long-term impact of false negatives on \ac{cert} operations remains unevaluated.
Additionally, we excluded domain-specific features(\eg, CVE identifiers, OWASP classifications) and foundational models~(\eg, Claude, DeepSeek, GPT) from our analysis.

Future work should address these limitations through expanded stability analysis, alternative distance metrics, systematic parameter sensitivity evaluation, and integration of domain-specific features while maintaining operational simplicity.
The exploration of multilingual capabilities and assessment of false negative impact on operational efficiency present additional research opportunities.

\section{Conclusion}
\label{clustering:sec:conclusion}

This work investigated the application of clustering algorithms for reducing information overload in \ac{cert} operations.
Through comprehensive evaluation of \num{196} cluster-embedding combinations across five datasets, we demonstrate that clustering can effectively reduce information processing requirements by over \num{90}~\% while maintaining semantic coherence.
However, optimal performance requires careful selection of clustering approaches based on specific data characteristics.

Deep clustering combined with \emph{stella} models demonstrates superior performance for structured security data~(\ac{sbr}, homogeneity \num{.84}), while partition-based clustering with \emph{NV-Embed-v2} excels for general advisory content~(\cysecalert{}, homogeneity \num{.61}).
The more complex \threatreport{} dataset requires parameter adjustment to achieve acceptable performance~(homogeneity~\num{.5}), highlighting the need for domain-specific tuning.
Runtime performance remains consistently efficient, with complete processing requiring approximately five minutes, potentially saving \acp{cert} over \num{3750} hours annually.

While our evaluation demonstrates clustering's potential for information overload reduction, several challenges remain.
Future work should address cluster separation optimization, systematic parameter tuning, and integration of domain-specific features while maintaining operational simplicity.
Despite these limitations, our findings suggest that clustering approaches, when properly configured, can significantly enhance \ac{cert} operational efficiency without compromising analytical integrity.

\begin{acks}
	This work was supported by the German Federal Ministry for Education and Research (BMBF) in the project CYWARN (13N15407)
	and
	German Federal Ministry of Education and Research and the Hessian Ministry of Higher Education, Research, Science and the Arts within their joint support of the National Research Center for Applied Cybersecurity ATHENE.

	The authors gratefully acknowledge the computing time provided to them on the high-performance computer Lichtenberg at the NHR Centers NHR4CES at TU Darmstadt. This is funded by the Federal Ministry of Education and Research, and the state governments participating on the basis of the resolutions of the GWK for national high performance computing at universities.
\end{acks}

\bibliographystyle{ACM-Reference-Format}
\bibliography{bibliography}

\section*{Appendix}
\label{sec:appendix}

The heatmaps show mean homogeneity scores for embedding-clustering combinations across four datasets.
Higher scores~(yellow) indicate better clustering: MSE~(\num{0.4}), \cysecalert{}~(\num{0.6}), SMS~(\num{0.8}), and \threatreport-60~(\num{0.2}--\num{0.5}).
Marginal means summarize model and algorithm performance. \medskip

\includegraphics[width=\linewidth]{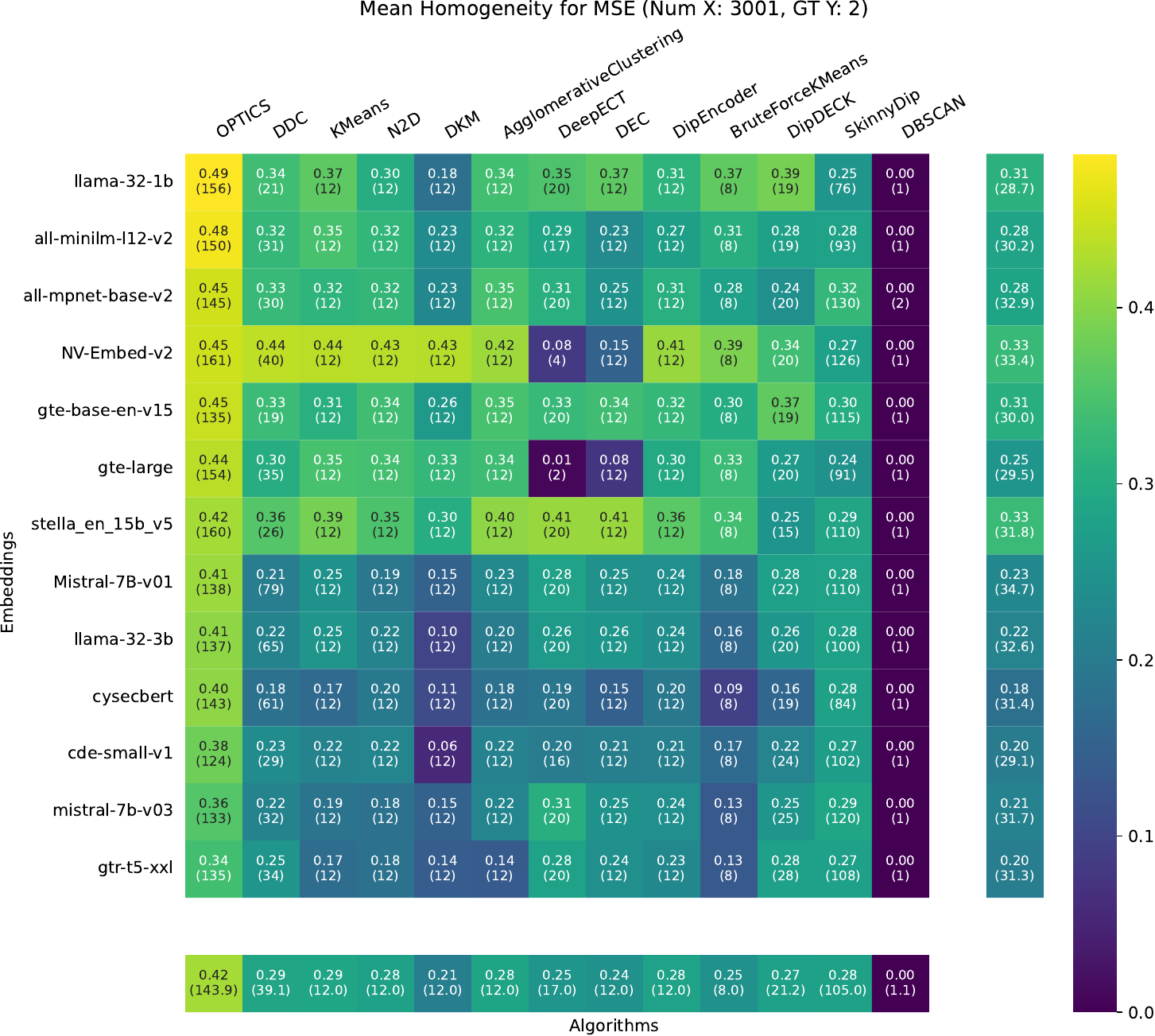}\medskip

\includegraphics[width=\linewidth]{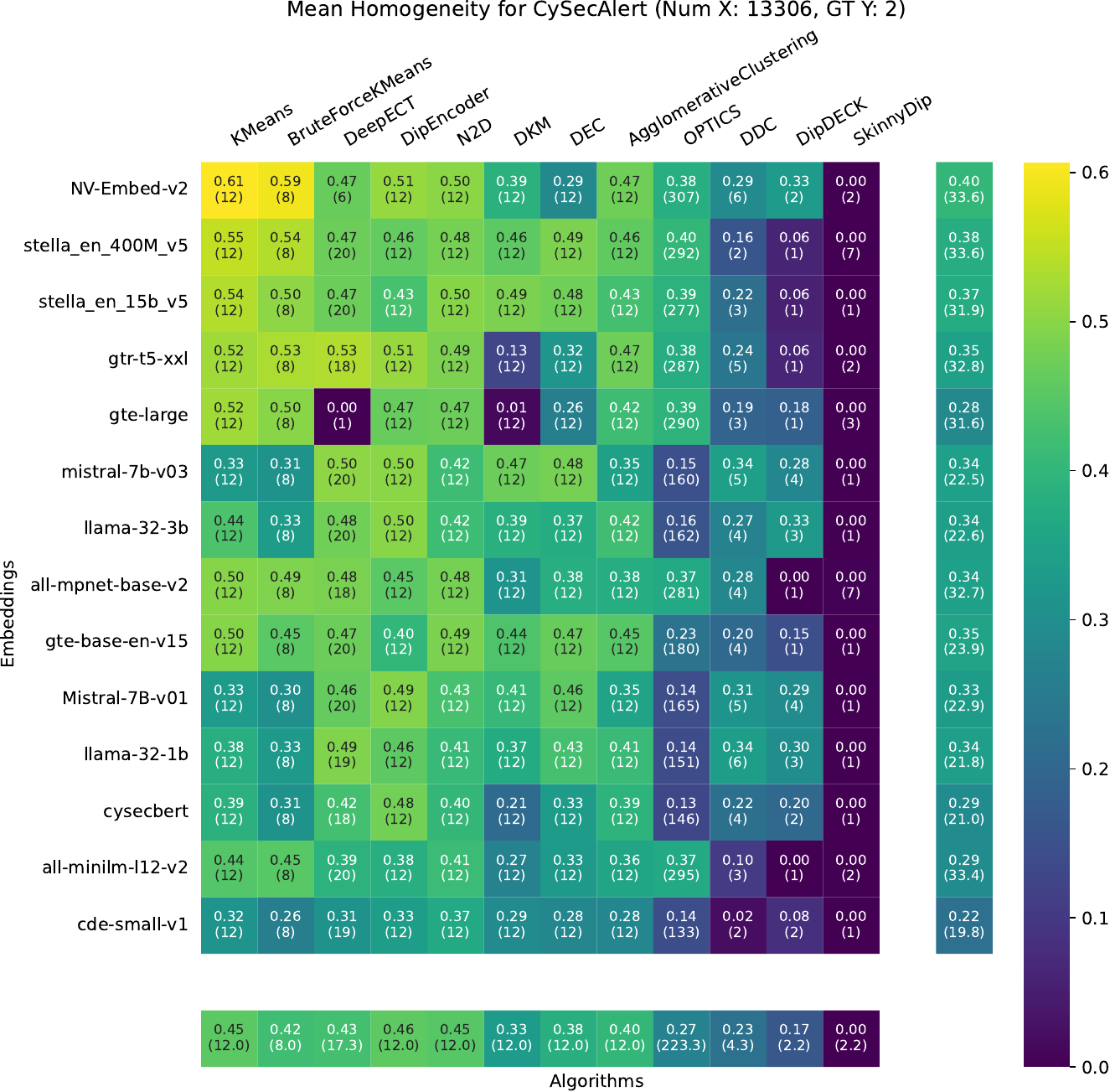}\medskip

\includegraphics[width=\linewidth]{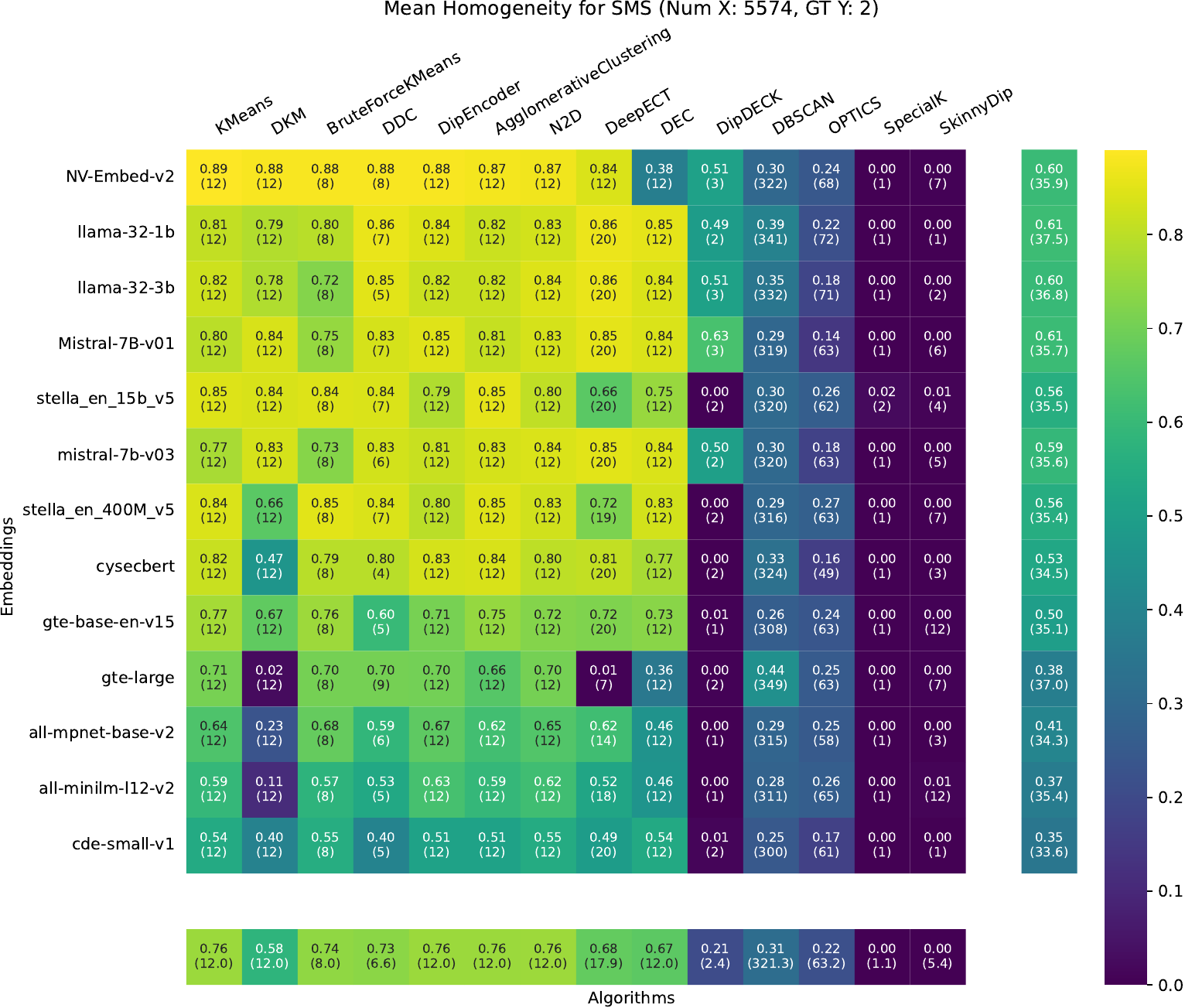}\medskip

\includegraphics[width=\linewidth]{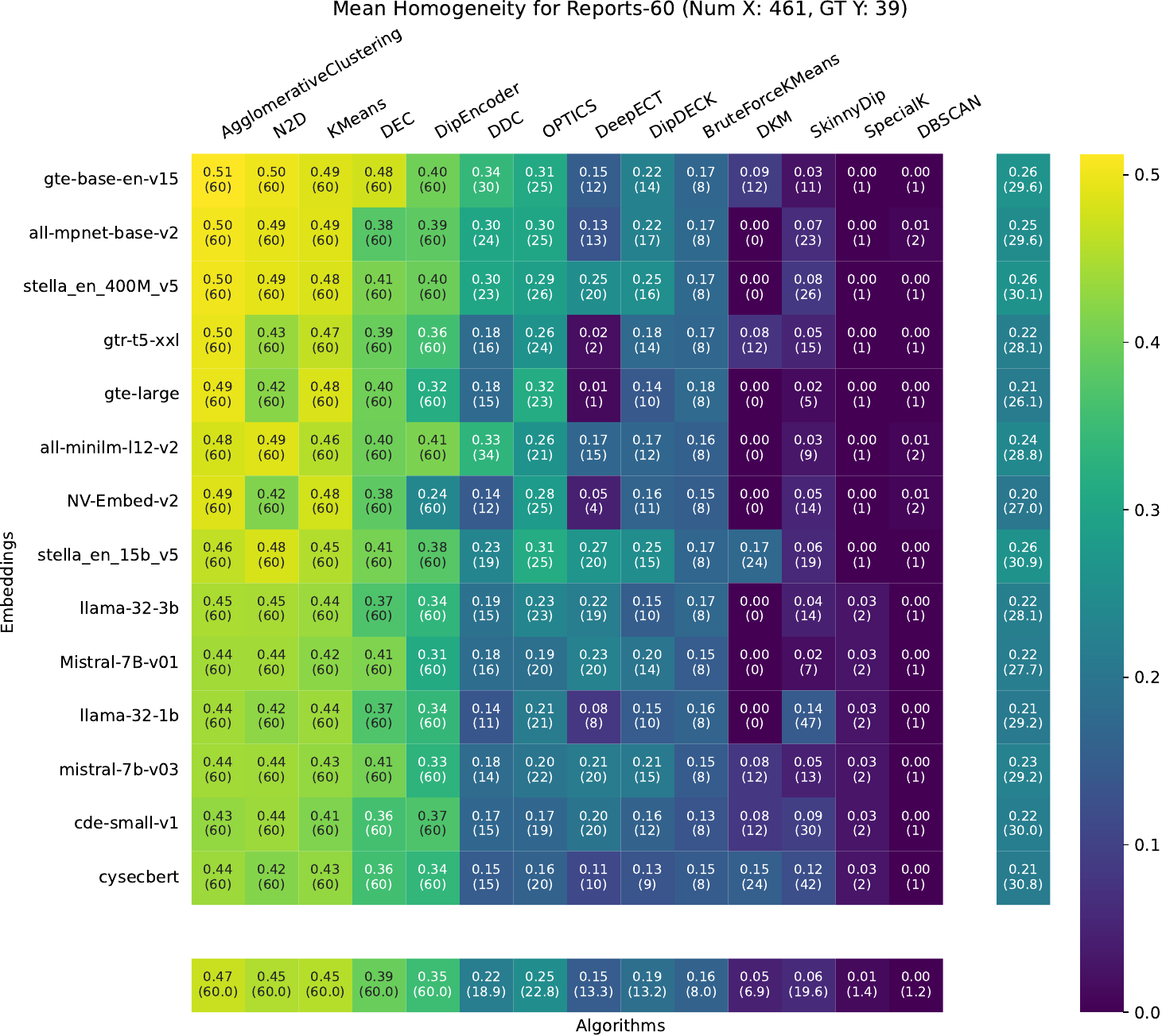}\medskip

\end{document}